# STATIC ANALYSIS FOR BIOLOGICAL SYSTEMS "BIOAMBIENTS"

A.Aziz Altowayan
aa10212w@pace.edu

**Abstract.** *In this paper, I present a summary on some works that utilized static analysis techniques for understanding biological systems. Control flow analysis, context dependent analysis, and other techniques were employed to investigate the properties of BioAmbients. In this summary report, I tried to introduce the ideas and explain the techniques used in the subject papers. This summary will highlight the biological concepts of BioAmbients.*

## 1. Introduction

Static analysis is a promising approach in the field of biology. In the work I investigated, they demonstrated how static analysis methods could help researchers to understand the dynamics of bio-systems. In 2003, the first attempt at utilizing this approach was introduced in work [2], where they exploited control flow analysis to track the contents of BioAmbients and the relations between them. The results provided fair over-approximation about the ambients' interactions. The main idea behind that work is to lift the current research practice in biology to a global view of biological systems that can be investigated at different levels of abstractions [1].

In fact, the complexity of biological systems constitutes a challenge for computer science [2]. Furthermore, analyzing the dynamics of these systems requires innovative methods to cope with the state-space explosion [7]. Recently and after the collected data about bio-components in the post-genomic era, modeling the dynamical behavior of biological systems is becoming an urgent need to biologists [2]. In the coming years, it is going to be a real challenge for computer science to apply a new paradigm that can make a shift from structure representation to behavior depiction yielding functional genomics [1].

Envision that we know what a car looks like, that we know all the details about its construction. Thus, we are able to simulate its structure. Moreover, we continue to receive more data about its organizational aspects. But on the other hand, we know nothing about how the car works. Thus, we cannot model the dynamics of its working process. This is the case in biomolecular systems. Biology has accumulated tremendous amounts of data about biological entities, but we know very little information on the functions of those entities [1]. The investigated papers are promising steps towards the definition of modeling environments for biologists to help them in the analysis of such complex systems.

### *When did it start?*

It is possible to model and simulate the dynamics of molecular systems using approaches based on process algebra[1] [2]. In 1994, the first work to use calculus for modeling biochemical systems was introduced [8] where they develop a minimal theory based on two abstractions from

1 Process algebras are a diverse family of related approaches for formally modeling concurrent systems [8]. It refers to behavior of a system [5].

chemistry. The theory is formulated using λ-calculus. However, expressions used in the biological society and computer science are different. As an attempt to bridge this gap, *bio-calculus* was proposed [9]. Bio-calculus is an expression system. It provides syntax that is similar to conventional expressions in biology and at the same time specifies information needed for simulation analysis. Also in 2003, a simple *core modeling language* for molecular biology was introduced [11]. In the same year, [3] proposed BioAmbients calculus *'motivated by the ambient calculus'* that could be used for representing the various aspects of molecular movements and localization.

The new paradigm to investigate biological systems is hypothesis driven, iterative, and global. It presents the possibility to investigate such complex systems at a different level of abstractions. The focus is no longer on the structure of biomolecular entities, but on the dynamics of behavior of these entities during their biological evolution [1]. Thus, we need some smart techniques to analyze the evolution of the biomolecular components. Static analysis techniques are good candidates for that purpose.

The next sections of this report are organized as follows. Section 2 gives an introduction to the biological concepts, explains compartments and their abstraction BioAmbient, and defines BioAmbient primitives. In section 3, I present a summary on the analysis techniques. Section 4 discusses some related works. Finally, I concluded this report in section 5. The sections 2 and 3 are mainly based on work done in [1], [2], and [3].

## 2. Biological Concept

Biomolecular systems consist of networks of proteins. Proteins are large molecules consisting of one or more chains of amino acids. Proteins perform a vast array of functions within living cells, and transport molecules from one location to another [14]. A biological compartment, *a closed part within a cell surrounded by a membrane*, is the key element in the structure of such systems [3]. Compartments are complex entities. We can look at biological compartments from 2 different perspectives, physical and functional. The physical compartments define the contents bounded within a membrane, i.e. nucleus, cytoplasm, and organelles (*see* Figure 1). On the other hand, the functional compartments define the behavior of the contents, i.e. activation/inactivation, targeting, and signaling [12].

Compartments play a crucial role in the functioning of biomolecular systems. The hierarchical organizing of compartments enables a molecule to perform its function. In the work [3], they presented BioAmbients primitives as abstraction for the biological compartments.

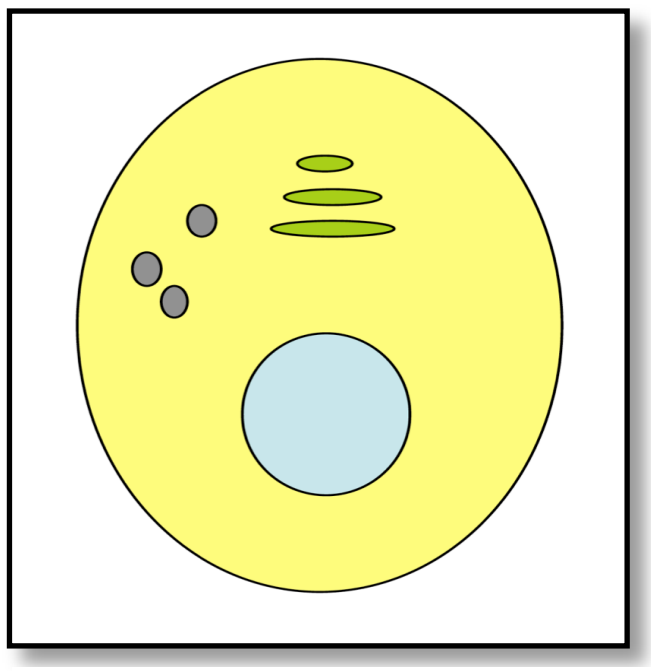

*Figure 1: Compartment of cells*

## *Compartments*

Most bio-entities may reside either within or outside some given compartments. Therefore, compartments introduce the notion of location. They can move or relocate from one place to another. Based on that, two types of movements are observed for compartment entities. *Movement between compartment*s is a molecule moving across a membrane (enter/exit), and *compartment movement*, a compartment joining another compartment, which is 2 compartments become one (merge).

The cellular compartments are diverse and have more than one type. Therefore, dealing with abstraction will facilitate the analysis task to capture the common properties. Using π-calculus process algebra, an abstraction for biomolecular systems was previously developed. In [3], the authors have extended this abstraction to handle compartments. The result of this work is presented as BioAmbients calculus, for representing various aspects of molecular localization and compartmentalization.

## *Compartments abstracted as BioAmbients*[2]

To facilitate the mapping of biological compartments, ambient calculus was modified into BioAmbients. Ambients are bounded places where computations take place and they have properties. Each ambient harbors a collection of processes running and residing in it. A process controls the ambient behavior, by instructing it where to move. An ambient can move from one location to another, as a whole with all its component processes and sub-ambients. There are boundaries surrounding the ambient and defining what is inside and what is outside. Ambients might be nested within other ambients (i.e. collections of sub-ambients). They are nameless, but to improve the readability and talk about its contents, annotated names will be used to distinguish between the various syntactic occurrences of ambients.

Ambients' ability to move defines what is called capabilities. Capabilities, or movements, can change the ambient hierarchy by allowing ambient entry, exit, or mergence. In movements, reactions are synchronous. That is, the subject and the object must agree on the reaction to happen. Therefore, capabilities are synchronized in pairs, using named channels. Ambients have three types of capabilities synced in pairs: *enter/exit* pair, *exit/expel* pair, and *merge+/merge-* pair.

The boundaries restrict the communications of ambients' processes. Information flow of ambients' movements is distinguished in three types of communication directions. *Local communication,* that is between 2 processes residing in the same ambient; *s2s communication,* which is between 2 processes residing in two sibling ambients; and *p2c/c2p communication*, that is between a process in a parent ambient and a process in its immediate child. The latter is asymmetric, depending on the location of the sender and receiver, whether information flows from the parent to the child or the other way. The communication directions are independent of the channel's identity.

---

2 This is based on work done in [3]

The abstraction is built through three universal steps: organization of the real-world domain “to capture the essential properties of biomolecular compartments,” selection (or development) of the mathematical domain “to describe the organization of BioAmbients,” and designing the abstraction between the two “to describe the application of the BioAmbients abstraction” [3]. The outcome defines the essential primitives to describe BioAmbients mobility and communication mentioned earlier (*see* Figure 2).

| | | | |
|---|---|---|---|
| $n, m, p$ | | | names |
| $\pi$ | $\stackrel{\text{def}}{=}$ | | Actions |
| | | $\$n!\{m\}$ | Output action |
| | | $\$n?\{m\}$ | Input action |
| $\$$ | $\stackrel{\text{def}}{=}$ | | Directions |
| | | $local$ | Intra-ambient |
| | | $s2s$ | Inter-siblings |
| | | $p2c$ | Parent to child |
| | | $c2p$ | Child to parent |

| | | | |
|---|---|---|---|
| $M, N$ | $\stackrel{\text{def}}{=}$ | | Capabilities |
| | | $enter\ n$ | Synch entry |
| | | $accept\ n$ | Synch accept |
| | | $exit\ n$ | Synch exit |
| | | $expel\ n$ | Synch expel |
| | | $merge+\ n$ | Synch merge with |
| | | $merge-\ n$ | Synch merge into |
| $P, Q$ | $\stackrel{\text{def}}{=}$ | | Processes |
| | | $(\mathsf{new}\ n)P$ | Restriction |
| | | $P \mid Q$ | Composition |
| | | $!P$ | Replication |
| | | $[P]$ | Ambient (membrane) |
| | | $\pi.P$ | Communication prefix |
| | | $M.P$ | Capability prefix |
| | | $\sum_{i \in I} \pi_i.P_i$ | Communication Choice |
| | | $\sum_{i \in I} M_i.P_i$ | Capability Choice |

*Figure 2: Mobility and communication primitives*

From a biological perspective, the given primitives can be used to describe ambients’ movements (*see* figure 3) and communication directions (*see* figure 4) [3]. In the communication primitives, BioAmbients use names as channels, and, thus, only names can be exchanged as a communication result.

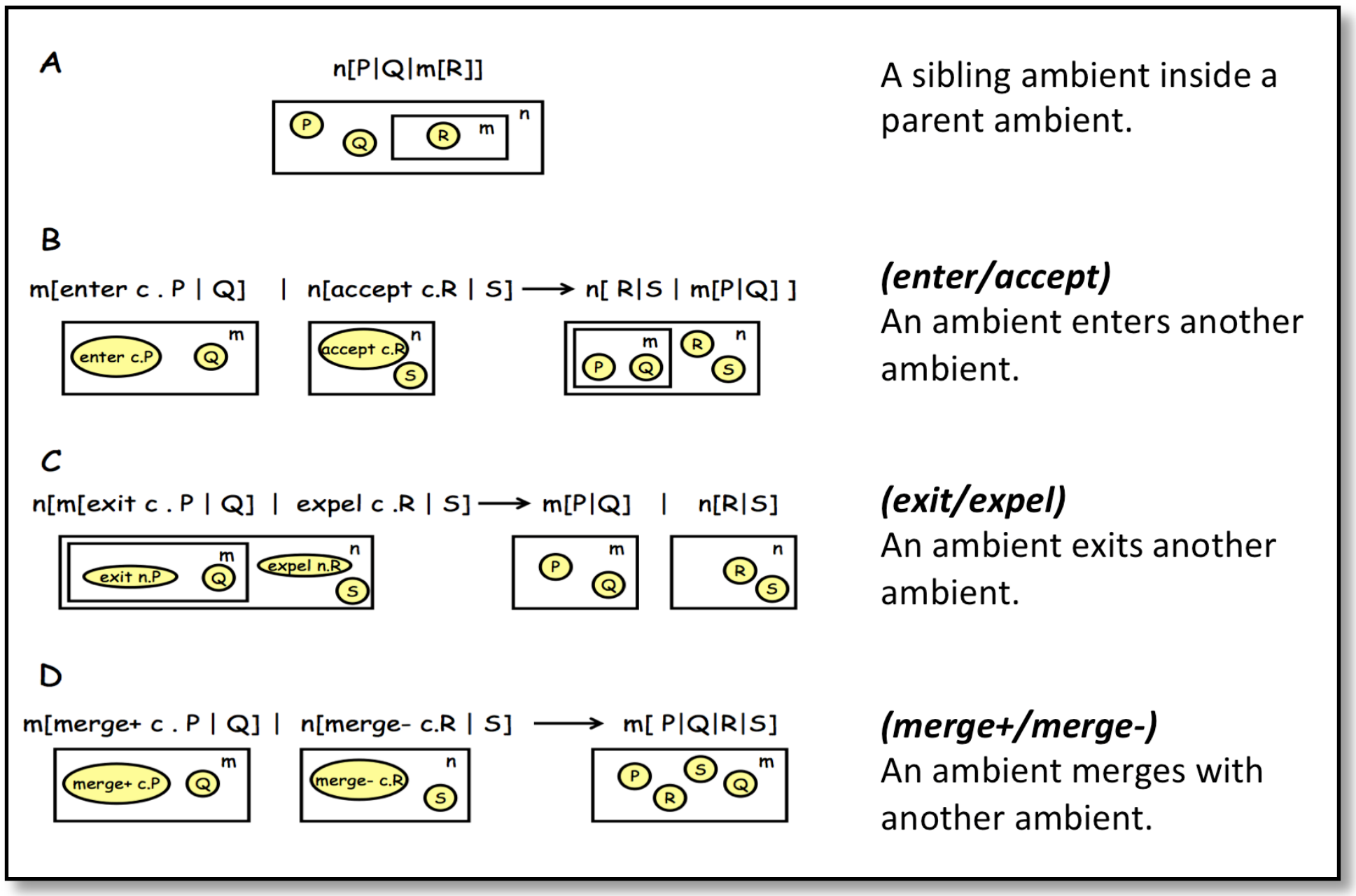


*Figure 3: Ambient movements: enter/accept 'B', exit/expel 'C', and merge+/merge- 'D'*

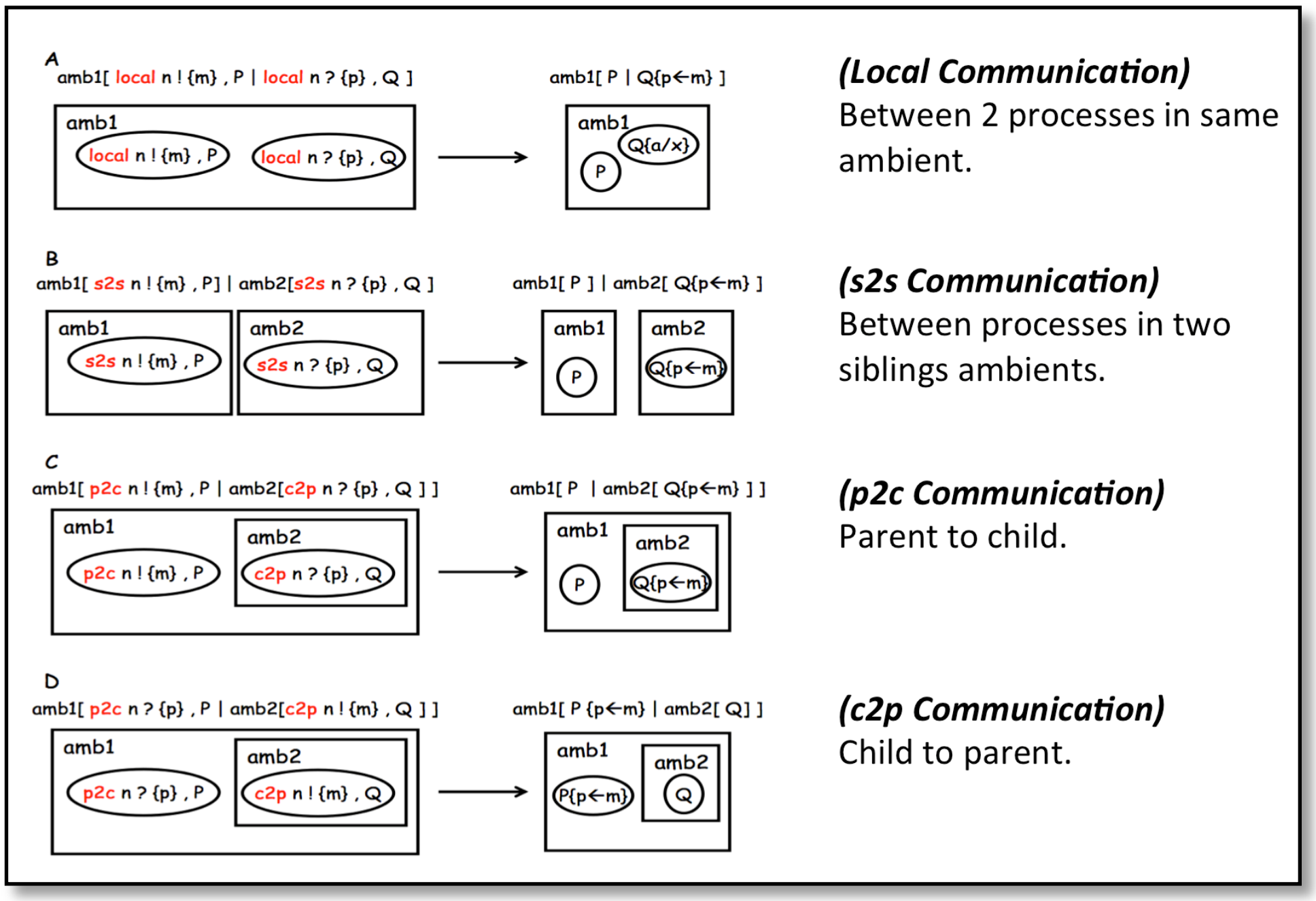


*Figure 4: Communication Directions 'note that p2c/c2p is asymmetric with 2 reduction rules'*

## 3. Static Analysis

Historically, static analysis techniques have been developed and used in the context of optimizing compilers as the classical application domain of the analysis. Here, a software code is analyzed to investigate the properties of a program without actually executing the program. In another related application, static analysis has also been used in software quality assurance. The techniques of static-program analysis offer safe predictions and computable approximations to the set of values or behaviors arising dynamically at run-time [4].

For a particular property of interest, the extracted static analysis information is guaranteed to be correct. However, it is important to note that it is impossible to obtain the exact information about the property's dynamic behavior. Therefore, static analysis introduces the notion of approximation. Approximation here can be crucial. For instance, when we have an over-approximation to the exact behavior, we can guarantee that certain events will never occur. On the other hand, when we have an under-approximation, we can make sure that certain events will indeed occur. Results that are neither over- nor under-approximations are uninteresting because we could not interpret them meaningfully (*see* figure 5) [1][4].

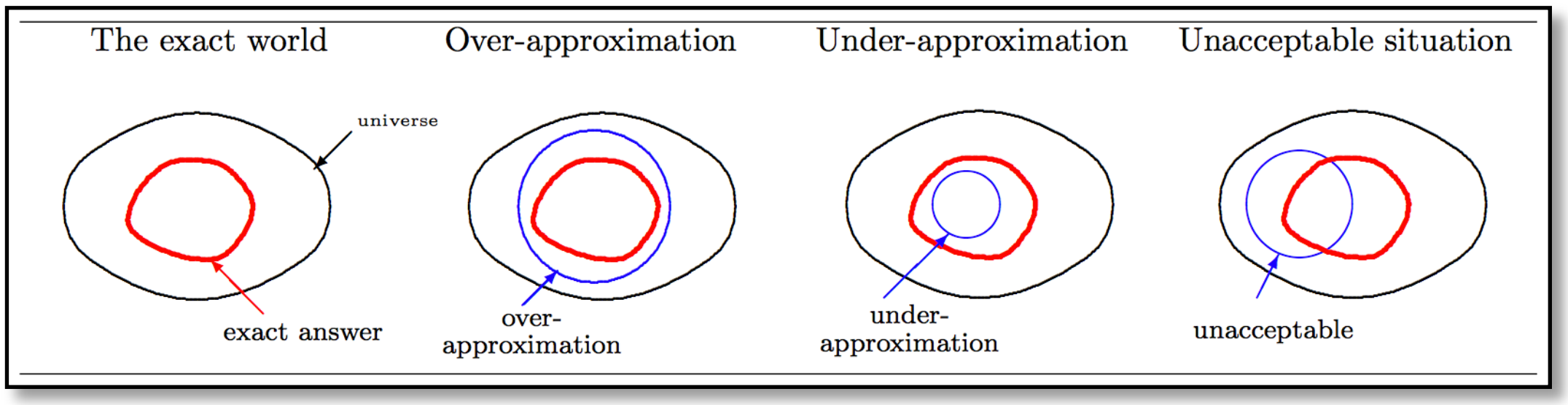


*Figure 5: Static Analysis is an approximative approach*

However, previous works have shown that static analysis approaches can be applied to handle a variety of application domains. The necessary constructs, for such variant domains, include mobility and communication primitives as in the π-calculus, mobile ambients, and boxed ambients [2]. In this report, we investigate the biological application domain as the subject domain of the analysis, where the approach is dedicated for analyzing molecular processes specified in BioAmbients.

In other works [11], *dynamic analysis* has been used for analyzing the molecular systems by relying on a transition-system representation to explore the properties of biomolecular processes. However, the dynamic analysis approach is limited because of the huge size of the representation. In transition systems, the size of the system is exponential to the program representing the behavior. Hence when the size of representation is too big, static analysis is a much more efficient alternative. To use static analysis in modeling a system, we only need the text of the program to infer suitable properties about the system's behavior [2].

## *Control Flow Analysis for BioAmbients*[3]

The primitives introduced [3] in compartments abstraction have been utilized for the design of static analysis to investigate the processes specified in BioAmbients. Traditionally in code analysis, control flow analysis implicitly refers to determining the receiver(s) of function or method calls and is expressed in Control-Flow Graph [8]. But in our case, the aim of control flow analysis is to keep track of the contents of ambients and the binding names between them (*see* figure 6). Control Flow Analysis is exploited to decode the bindings of variables induced by communications and to build a relation of the ambients [2].

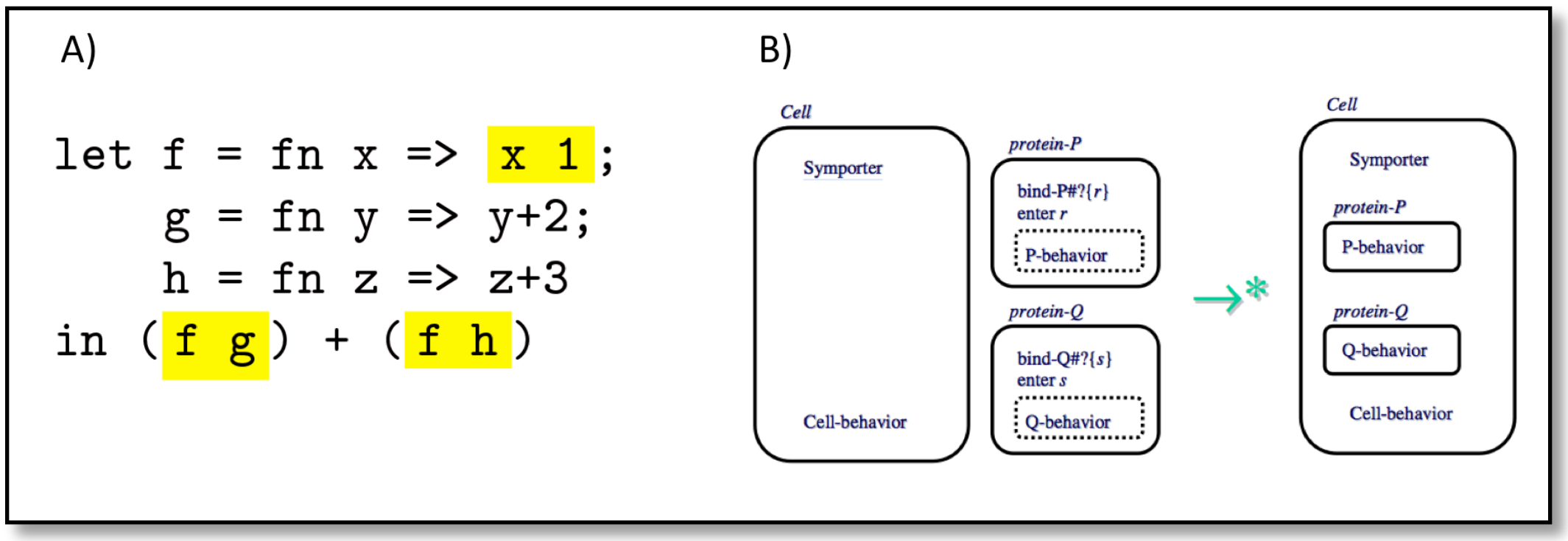


*Figure 6: Control Flow Analysis in Classical Domain Context 'A' and in Biological Domain Context 'B'*

The subjects of the analysis are the ambients' processes and their capabilities. Capabilities govern the behavior of processes. Influenced by these capabilities, ambients' processes communicate with one another to implement the various movements' actions and reactions discussed earlier. The syntax of BioAmbients is given in figure 7, where *P*: Processes and *M*: Capabilities. The syntax of processes and capabilities is adapted based on the mobility and communication primitives provided earlier.

Since ambients are nameless entities, they are annotated to distinguish between the various syntactic occurrences. The given syntax refers to ambients by their identity as in $[P]\,\mu$, where $\mu$ denotes the identity of ambient $[P]$. The movement capabilities are based on the *subject* and the *object*; containing capabilities share the same name. Therefore, the syntactic grammar rules reflect each capability as a parallel composition of two processes.

---

3 This is based on work done in [2]

| | |
|---|---|
| $P ::= \mathsf{0}$ | inactive process |
| $\mid\ (n)P$ | binding box for the constant $n$ |
| $\mid\ [P]^{\mu}$ | ambient $P$ with the role $\mu$ |
| $\mid\ M.P$ | prefixing with capability $M$ |
| $\mid\ P \mid P'$ | parallel processes |
| $\mid\ P + P'$ | non-deterministic external choice |
| $\mid\ \mathsf{rec}\, X.\, P$ | recursive process (*i.e.* $X = P$) |
| $\mid\ X$ | process identifier |
| $M ::= \mathsf{enter}\ n \mid \mathsf{accept}\ n$ | enter movement |
| $\mid\ \mathsf{exit}\ n \mid \mathsf{expel}\ n$ | exit movement |
| $\mid\ \mathsf{merge}\text{–}\ n \mid \mathsf{merge}+\ n$ | merge movement |
| $\mid\ n!\{m\} \mid n?\{p\}$ | local output and input binding the variable $p$ |
| $\mid\ n_!\{m\} \mid n\hat{}?\{p\}$ | parent to child output and input binding the variable $p$ |
| $\mid\ n\hat{}!\{m\} \mid n_?\{p\}$ | child to parent output and input binding the variable $p$ |
| $\mid\ n\#!\{m\} \mid n\#?\{p\}$ | sibling output and input binding the variable $p$ |

*Figure 7: Syntax of BioAmbients*

Ambients' communication yields the exchanging of names. Hence, names are subject to alpha-renaming because they cannot carry information in the analysis. Therefore, canonical names shall be used, i.e. each name $n$ has a canonical name written $\lfloor n \rfloor$. It is assumed that the canonical names are preserved under alpha-renaming. In the developed analysis, the approximation of ambients' contents and their bindings are represented as follows:

- The **contents** of an ambient is approximated as:

$$\mathcal{I} \subseteq \textbf{Ambient} \times (\textbf{Ambient} \cup \textbf{Cap})$$

*Where,*
**Ambient:** The set of ambient identities 'finite'.
**Cap:** The set of canonical capabilities.

This means, an ambient may contain other ambient as well as capabilities. So when we use $u \in \mathcal{I}(\mu)$, this means that $\mu$ may contain $u$. The analysis here is affected by *movement capabilities.*

- The relevant name **bindings** 'Relations' is approximated as:

$$\mathcal{R} \subseteq \textbf{Name} \times \textbf{Name}$$

*Where,*
**Name:** The set of canonical names.

So when $v' \in \mathcal{R}(v)$, it will mean that $v$ may take on the value $v'$. That means $v'$ will be the canonical name of the name being transmitted. The analysis here is affected by *communication capabilities.*

From the analysis perspective, the transition relation of ambients' capabilities can be described using BioAmbient syntax as shown in figure 8. The analysis transition demonstrates each movement capability happening between processes of ambients $\mu_1$ and ambient $\mu_2$ "*entry/accept*, *exit/expel*, and *merge+/merge-*" as in (figure8, A) [4]. The analysis transition for communication is also demonstrated for each direction *local, s2s, p2c/c2p* (figure 8, B).

**A)** *Ambients Movements:*

$$[(\text{enter } n.P + P') \mid P'']^{\mu_1} \mid [(\text{accept } n.Q + Q') \mid Q'']^{\mu_2} \rightarrow [[P \mid P'']^{\mu_1} \mid Q \mid Q'']^{\mu_2}$$

$$[[(\text{exit } n.P + P') \mid P'']^{\mu_1} \mid (\text{expel } n.Q + Q') \mid Q'']^{\mu_2} \rightarrow [P \mid P'']^{\mu_1} \mid [Q \mid Q'']^{\mu_2}$$

$$[(\text{merge}+ \ n.P + P') \mid P'']^{\mu_1} \mid [(\text{merge}- \ n.Q + Q') \mid Q'']^{\mu_2} \rightarrow [P \mid P'' \mid Q \mid Q'']^{\mu_1}$$

**B)** *Ambients Communication*:

$$(n!\{m\}.P + P') \mid (n?\{p\}.Q + Q') \rightarrow P \mid Q[m/p]$$

$$(n_!\{m\}.P + P') \mid [(n\hat{}?\{p\}.Q + Q') \mid Q'']^{\mu} \rightarrow P \mid [Q[m/p] \mid Q'']^{\mu}$$

$$[(n\hat{}!\{m\}.P + P') \mid P'']^{\mu} \mid (n_?\{p\}.Q + Q') \rightarrow [P \mid P'']^{\mu} \mid Q[m/p]$$

$$[(n\#!\{m\}.P + P') \mid P'']^{\mu_1} \mid [(n\#?\{p\}.Q + Q') \mid Q'']^{\mu_2} \rightarrow [P \mid P'']^{\mu_1} \mid [Q[m/p] \mid Q'']^{\mu_2}$$

*Figure 8: (Transition relation P → P') movements in 'A', and communication directions in 'B'*

The execution of processes transition can be described as in figure 9. Note that the analysis take into account that if $P$ evolves into $P'$ in a number of steps, i.e. $P \rightarrow \ldots P'$, then both processes are considered as an over-approximated result for the analysis. However, this is a non-deterministic choice during the analysis, i.e. $P + P'$ (*see* figure 8).

*Execution in context:*

$$\frac{P \rightarrow Q}{(n)P \rightarrow (n)Q} \qquad \frac{P \rightarrow Q}{[P]^{\mu} \rightarrow [Q]^{\mu}} \qquad \frac{P \rightarrow Q}{P \mid R \rightarrow Q \mid R} \qquad \frac{P[\text{rec } X.P/X] \rightarrow Q}{\text{rec } X.P \rightarrow Q}$$

$$\frac{P \equiv P' \quad P' \rightarrow Q' \quad Q' \equiv Q}{P \rightarrow Q}$$

*Figure 9: (Transition relation P → P') during the execution.*

4 $n$.X stands for the channel $n$ used by a process X.

For the analysis judgments we use the forms:

$(\mathcal{I}, \mathcal{R}) \models^{\star} P$ (for processes)
*and*,
$(\mathcal{I}, \mathcal{R}) \models^{\star} M$ (for capabilities) *where*, $\star \in$ **Ambient**

Two stages are used to specify the analysis. First is to make sure that $\mathcal{I}$ and $\mathcal{R}$ describe the same initial process, this is defined in (figure 10) below for both processes and capabilities [5].

**A) Analysis of Processes**

$$(\mathcal{I}, \mathcal{R}) \models^{\star} P$$

| | | |
|---|---|---|
| $(\mathcal{I}, \mathcal{R}) \models^{\star} 0$ | iff | true |
| $(\mathcal{I}, \mathcal{R}) \models^{\star} (n)P$ | iff | $\lfloor n \rfloor \in \mathcal{R}(\lfloor n \rfloor) \wedge (\mathcal{I}, \mathcal{R}) \models^{\star} P$ |
| $(\mathcal{I}, \mathcal{R}) \models^{\star} [P]^{\mu}$ | iff | $\mu \in \mathcal{I}(\star) \wedge (\mathcal{I}, \mathcal{R}) \models^{\mu} P$ |
| $(\mathcal{I}, \mathcal{R}) \models^{\star} M.P$ | iff | $(\mathcal{I}, \mathcal{R}) \models^{\star} M \wedge (\mathcal{I}, \mathcal{R}) \models^{\star} P$ |
| $(\mathcal{I}, \mathcal{R}) \models^{\star} P \mid P'$ | iff | $(\mathcal{I}, \mathcal{R}) \models^{\star} P \wedge (\mathcal{I}, \mathcal{R}) \models^{\star} P'$ |
| $(\mathcal{I}, \mathcal{R}) \models^{\star} P + P'$ | iff | $(\mathcal{I}, \mathcal{R}) \models^{\star} P \wedge (\mathcal{I}, \mathcal{R}) \models^{\star} P'$ |
| $(\mathcal{I}, \mathcal{R}) \models^{\star} \text{rec}\, X.\, P$ | iff | $(\mathcal{I}, \mathcal{R}) \models^{\star} P$ |
| $(\mathcal{I}, \mathcal{R}) \models^{\star} X$ | iff | true |

**B) Analysis of Capabilities**

$$(\mathcal{I}, \mathcal{R}) \models^{\star} M$$

| | | |
|---|---|---|
| $(\mathcal{I}, \mathcal{R}) \models^{\star} \text{enter}\ n$ | iff | $\forall \nu_n : \nu_n \in \mathcal{R}(\lfloor n \rfloor) \Rightarrow \text{enter}\ \nu_n \in \mathcal{I}(\star)$ |
| $(\mathcal{I}, \mathcal{R}) \models^{\star} \text{accept}\ n$ | iff | $\forall \nu_n : \nu_n \in \mathcal{R}(\lfloor n \rfloor) \Rightarrow \text{accept}\ \nu_n \in \mathcal{I}(\star)$ |
| $(\mathcal{I}, \mathcal{R}) \models^{\star} \text{exit}\ n$ | iff | $\forall \nu_n : \nu_n \in \mathcal{R}(\lfloor n \rfloor) \Rightarrow \text{exit}\ \nu_n \in \mathcal{I}(\star)$ |
| $(\mathcal{I}, \mathcal{R}) \models^{\star} \text{expel}\ n$ | iff | $\forall \nu_n : \nu_n \in \mathcal{R}(\lfloor n \rfloor) \Rightarrow \text{expel}\ \nu_n \in \mathcal{I}(\star)$ |
| $(\mathcal{I}, \mathcal{R}) \models^{\star} \text{merge+}\ n$ | iff | $\forall \nu_n : \nu_n \in \mathcal{R}(\lfloor n \rfloor) \Rightarrow \text{merge+}\ \nu_n \in \mathcal{I}(\star)$ |
| $(\mathcal{I}, \mathcal{R}) \models^{\star} \text{merge–}\ n$ | iff | $\forall \nu_n : \nu_n \in \mathcal{R}(\lfloor n \rfloor) \Rightarrow \text{merge–}\ \nu_n \in \mathcal{I}(\star)$ |
| $(\mathcal{I}, \mathcal{R}) \models^{\star} n!\{m\}$ | iff | $\forall \nu_n, \nu_m : \nu_n \in \mathcal{R}(\lfloor n \rfloor) \wedge \nu_m \in \mathcal{R}(\lfloor m \rfloor) \Rightarrow \nu_n!\{\nu_m\} \in \mathcal{I}(\star)$ |
| $(\mathcal{I}, \mathcal{R}) \models^{\star} n?\{p\}$ | iff | $\forall \nu_n : \nu_n \in \mathcal{R}(\lfloor n \rfloor) \Rightarrow \nu_n?\{\lfloor p \rfloor\} \in \mathcal{I}(\star)$ |
| $(\mathcal{I}, \mathcal{R}) \models^{\star} n_!\{m\}$ | iff | $\forall \nu_n, \nu_m : \nu_n \in \mathcal{R}(\lfloor n \rfloor) \wedge \nu_m \in \mathcal{R}(\lfloor m \rfloor) \Rightarrow \nu_n_!\{\nu_m\} \in \mathcal{I}(\star)$ |
| $(\mathcal{I}, \mathcal{R}) \models^{\star} n\hat{}?\{p\}$ | iff | $\forall \nu_n : \nu_n \in \mathcal{R}(\lfloor n \rfloor) \Rightarrow \nu_n\hat{}?\{\lfloor p \rfloor\} \in \mathcal{I}(\star)$ |
| $(\mathcal{I}, \mathcal{R}) \models^{\star} n\hat{}!\{m\}$ | iff | $\forall \nu_n, \nu_m : \nu_n \in \mathcal{R}(\lfloor n \rfloor) \wedge \nu_m \in \mathcal{R}(\lfloor m \rfloor) \Rightarrow \nu_n\hat{}!\{\nu_m\} \in \mathcal{I}(\star)$ |
| $(\mathcal{I}, \mathcal{R}) \models^{\star} n_?\{p\}$ | iff | $\forall \nu_n : \nu_n \in \mathcal{R}(\lfloor n \rfloor) \Rightarrow \nu_n_?\{\lfloor p \rfloor\} \in \mathcal{I}(\star)$ |
| $(\mathcal{I}, \mathcal{R}) \models^{\star} n\#!\{m\}$ | iff | $\forall \nu_n, \nu_m : \nu_n \in \mathcal{R}(\lfloor n \rfloor) \wedge \nu_m \in \mathcal{R}(\lfloor m \rfloor) \Rightarrow \nu_n\#!\{\nu_m\} \in \mathcal{I}(\star)$ |
| $(\mathcal{I}, \mathcal{R}) \models^{\star} n\#?\{p\}$ | iff | $\forall \nu_n : \nu_n \in \mathcal{R}(\lfloor n \rfloor) \Rightarrow \nu_n\#?\{\lfloor p \rfloor\} \in \mathcal{I}(\star)$ |

*Figure 10: Judgment of the analysis: 'A' for processes, and 'B' for capabilities.*

Second is to make sure that $\mathcal{I}$ and $\mathcal{R}$ take the dynamics of the process into account, this is formulated by the closure conditions as in (figure 11).

---

5 $\mathcal{I}$: Approximation of the ambients' contents, and $\mathcal{R}$: Approximation of name bindings.

**Closure condition on $\mathcal{J}$ and $\mathcal{R}$:**

$$
\begin{array}{lll}
\text{Enter/accept:} & \forall \mu, \mu_1, \mu_2, \nu_n : & \mathsf{enter}\ \nu_n \in \mathcal{I}(\mu_1) \wedge \mu_1 \in \mathcal{I}(\mu) \wedge \\
& & \mathsf{accept}\ \nu_n \in \mathcal{I}(\mu_2) \wedge \mu_2 \in \mathcal{I}(\mu) \\
& & \Rightarrow \mu_1 \in \mathcal{I}(\mu_2) \\
\text{Exit/expel :} & \forall \mu, \mu_1, \mu_2, \nu_n : & \mathsf{exit}\ \nu_n \in \mathcal{I}(\mu_1) \wedge \mu_1 \in \mathcal{I}(\mu_2) \wedge \\
& & \mathsf{expel}\ \nu_n \in \mathcal{I}(\mu_2) \wedge \mu_2 \in \mathcal{I}(\mu) \\
& & \Rightarrow \mu_1 \in \mathcal{I}(\mu) \\
\text{Merge:} & \forall \mu, \mu_1, \mu_2, \nu_n : & \mathsf{merge+}\ \nu_n \in \mathcal{I}(\mu_1) \wedge \mu_1 \in \mathcal{I}(\mu) \wedge \\
& & \mathsf{merge-}\ \nu_n \in \mathcal{I}(\mu_2) \wedge \mu_2 \in \mathcal{I}(\mu) \\
& & \Rightarrow \forall \mu' : \mu' \in \mathcal{I}(\mu_2) \Rightarrow \mu' \in \mathcal{I}(\mu_1) \\
\text{To local:} & \forall \mu, \nu_m, \nu_p, \nu_n : & \nu_n!\{\nu_m\} \in \mathcal{I}(\mu) \wedge \\
& & \nu_n?\{\nu_p\} \in \mathcal{I}(\mu) \\
& & \Rightarrow \nu_m \in \mathcal{R}(\nu_p) \\
\text{To child:} & \forall \mu, \mu_c, \nu_m, \nu_p, \nu_n : & \nu_n_!\{\nu_m\} \in \mathcal{I}(\mu) \wedge \\
& & \nu_n\hat{}?\{\nu_p\} \in \mathcal{I}(\mu_c) \wedge \mu_c \in \mathcal{I}(\mu) \\
& & \Rightarrow \nu_m \in \mathcal{R}(\nu_p) \\
\text{To parent:} & \forall \mu, \mu_c, \nu_m, \nu_p, \nu_n : & \nu_n\hat{}!\{\nu_m\} \in \mathcal{I}(\mu_c) \wedge \mu_c \in \mathcal{I}(\mu) \wedge \\
& & \nu_n_?\{\nu_p\} \in \mathcal{I}(\mu) \\
& & \Rightarrow \nu_m \in \mathcal{R}(\nu_p) \\
\text{To sibling:} & \forall \mu, \mu_1, \mu_2, \nu_m, \nu_p, \nu_n : & \nu_n\#!\{\nu_m\} \in \mathcal{I}(\mu_1) \wedge \mu_1 \in \mathcal{I}(\mu) \wedge \\
& & \nu_n\#?\{\nu_p\} \in \mathcal{I}(\mu_2) \wedge \mu_2 \in \mathcal{I}(\mu) \\
& & \Rightarrow \nu_m \in \mathcal{R}(\nu_p)
\end{array}
$$

*Figure 11: Closure condition of the approximation*

Now, the analysis result can describe an over-approximation to the actual behavior of the process. This is described in the following **Theorem***;* this will take into account *all* the actual transition steps of the process:

*Assume*:

$$P \rightarrow Q, (\mathcal{J}, \mathcal{R}) \models_{\star} P \quad \textit{and},$$
$$\forall n \in \mathsf{fn}(P) \ : \ \lfloor n \rfloor\ \mathcal{R}(\lfloor n \rfloor).$$

*Then*,

$$(\mathcal{J}, \mathcal{R}) \models_{\star} Q$$

*Figure 12: Analysis Theorem*

*Where*,
$\mathsf{fn}(P)$: The set of free names in $P$

*Example on using the analysis*[6]

To illustrate the analysis approach and describe how ambients are written in code, we utilize BioAmbient syntax to write the following simple program as a BioAmbient process:

$$
\begin{aligned}
(c)(cell_1)(cell_2)(cell_3)\ \Big[\ & \mathsf{rec}\,X.\ (\mathsf{enter}\ cell_1.\ X + \mathsf{exit}\ cell_2.\ X + c\hat{}?\{x\}.\ \mathsf{expel}\ x.\ X) \\
& \mid [\mathsf{exit}\ cell_3.\ 0]^{D}\ \Big]^{mol} \\
& \mid [\mathsf{rec}\,X.\ (\mathsf{accept}\ cell_1.\ X + \mathsf{expel}\ cell_2.\ X + c_!\{cell_3\}.\ X)]^{cell}
\end{aligned}
$$

*where*,
"+" Choice between various actions
"." Sequencing of actions

The process contains three ambients *cell, mol,* and *D*, where *D* is inside *mol* and *cell* and *mol* represent the top-level. Ambients' capabilities govern the behavior of the process. In the written program, capabilities make use of the four names *c*, $cell_1$, $cell_2$, and $cell_3$ to implement the actions associated with movements. The ambients *cell* and *mol* consist of processes that can repeat themselves, where *D* has a process that can be executed only once. The process behavior is described as follows:

- *mol* executes **enter** $cell_1$ capability and *cell* executes **accept** $cell_1$ capability. Now, *mol* is inside *cell* as a result of this. Note that ambient *D* is inside *mol* so it will be moved inside *cell* as well.
- Two possible scenarios can happen after this capability is executed:
    a) *mol* executes **exit** $cell_2$ and *cell* executes **expel** $cell_2$; that will make *mol* leave *cell* then we will be back to the initial configuration. Thus, capabilities **enter/accept** and **exit/expel** can be repeated again resulting in *mol* moveing in and out *cell* multiple times.
    b) *cell* sends the name $cell_3$ on the channel *c* and *mol* receives a name on *c* and binds it to *x*. When this happens, capability **expel** *x* of *mol* actually becomes **expel** $cell_3$ and *D* will execute the corresponding **exit** $cell_3$; as a result *D* will leave *mol* and will become a subambient of *cell*.

Three types of static analysis techniques *(control flow analysis, context dependent analysis,* and *pathway analysis)* were applied to discover the process behavior without running the process. They all compute an over-approximation to the exact behavior of the process. If wondering why we need to try different techniques, it is because they vary in their precision and the computational cost depending on our analysis specification. However, in our example all the applied techniques compute an over-approximation, which describes the *father-son relationship* between ambients and capabilities.

---

6 This is based on work done in [1]

For instance, in control-flow analysis we ignore the order in which the capabilities are executed, and also we do not take the context into account. Thus the analysis takes its starting point as given in the initial father-son relationship (the black edges in Figure 13). The analysis imitates the semantics of the BioAmbients. It continues to provide information about the father-son relationship until no further information can be deduced. The analysis also keeps track of the potential name bindings, in order to give a proper treatment of the communication capabilities. In figure 13, the oval nodes represent ambients, square nodes represent capabilities, and a father-to-son relationship is represented with edges. The red edges represent the information added in this stage of the analysis – note: this information on the relationship keeps changing until the process is eventually terminated.

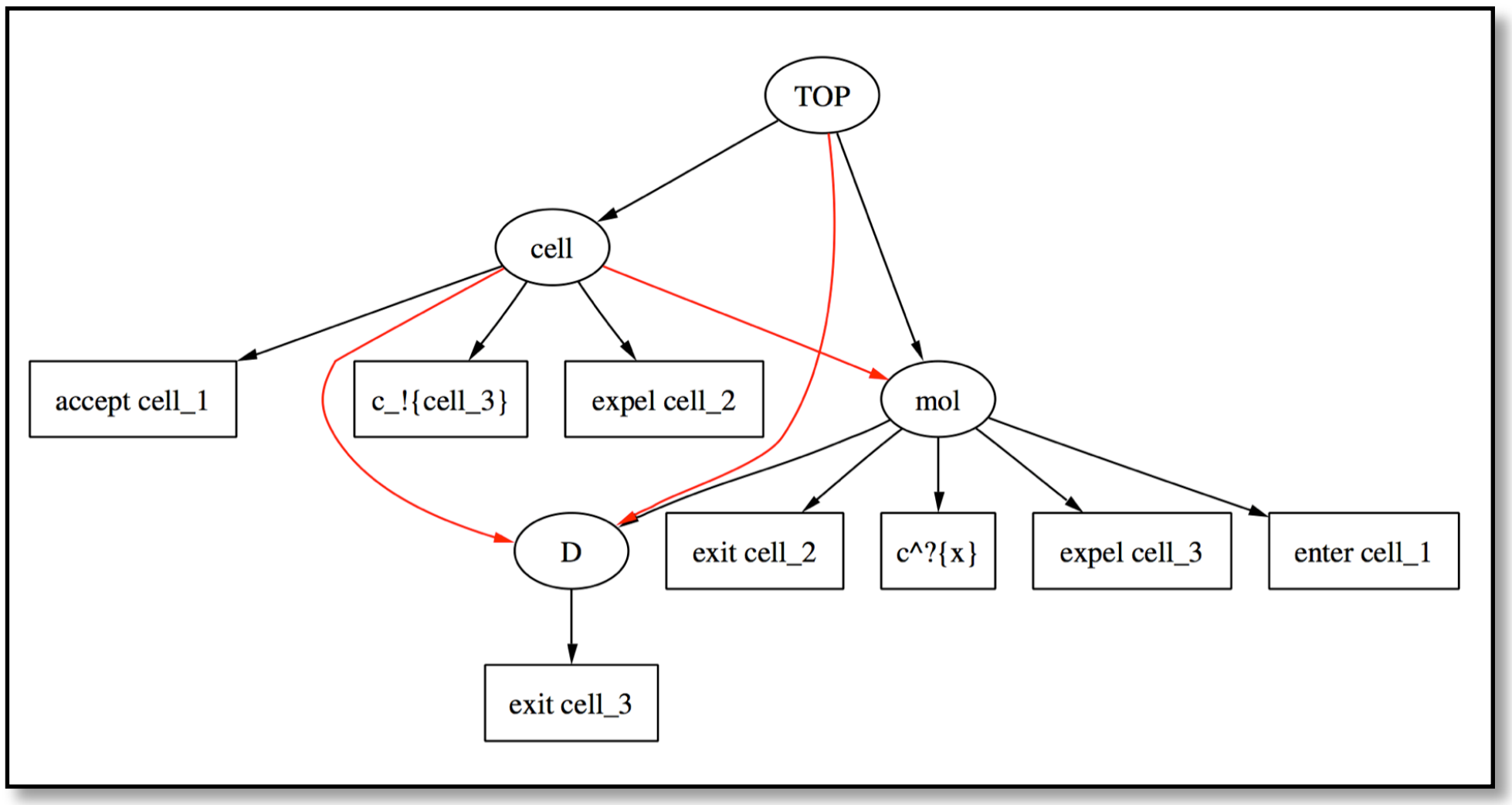

*Figure 13: Simple Control Flow Analysis result*

The analysis result is claimed to be indeed correct. It gives a precise over-approximation to the father-son relationship. What we learn from the result, in our example, is that *cell* will neither move into *mol* nor into *D*.

## 4. Related work

In 2004, Nielson [13] published an additional work for capturing the spatial structure of BioAmbients. The goal of spatial analysis is to extract an over-approximation of the possible hierarchical structures of ambients. It introduces the idea of *compatibility analysis* to identify the pairs of capabilities that *may* engage in a transition. In the following year 2005, Levi [6] used an abstract interpretation approach for modeling BioAmbients. A new occurrence counting analysis is proposed to i) record information about the number of occurrences of objects, and ii) keep more detail about the *possible* contents of ambients. These two novelties present a new dimension of complexity, since they specify the analysis for capturing *quantitative* information and *causality* aspects of BioAmbient interactions.

To some level of abstraction, it possible to regard biological systems as *concurrent systems*. BioAmbient Machine (BAM) is proposed [15] as a tool to implement a stochastic version of the BioAmbient calculus. BAM helps in understanding and reasoning the multiple interconnections between bioregulatory components.

Finally, the most recent attempt to exploiting static analysis for understanding biological systems was just published in 2012 [7]. The work proposes an original approach for deciding reachability based on *Process Hitting*. It uses a complementary abstract interpretation for analyzing the dynamics of biological regulatory networks (BRNs).

## 5. Conclusion

Understanding the dynamics of biological systems is the mutual goal of all works investigated in this report. Different static analysis techniques were dedicated to model and explore the interaction mechanisms in biomolecular components. Some works have established the feasibility of the approach, as in the case of control flow analysis [2].

Finally, such a new paradigm of interdisciplinary approach is useful for both computer science and biology. For biologists, it gives further insight into understanding the biological components, which should definitely lead to predictive and preventive medicine. On computer science side, "*Once the complete behavior of cells is well understood, we can think of programming biological entities*." [1]

This perspective, for programming the biological components, might sounds somehow unreachable in the near future. However, L. Cardelli [16], from Microsoft research in joint with Weizmann Institute of science, has proposed an emerging class of languages for describing biological systems, which possibly could pave the way for programming a living cell; it is called Bioware languages. In addition, F. Nielson [2] is currently investigating a methodology to code biological systems in BioAmbients.